# High-pressure x-ray diffraction study of bulk and nanocrystalline PbMoO$_4$

D. Errandonea[1,†,*], D. Santamaria-Perez[2,*], V. Grover[3], S. N. Achary[3], and A. K. Tyagi[3]

[1] Departamento de Física Aplicada-ICMUV, Universitat de València,

Edificio de Investigación, c/Dr. Moliner 50, 46100 Burjassot (Valencia), Spain.

[2] Departamento de Química Física I, Universidad Complutense de Madrid, Avda.

Complutense s/n, 28040 Madrid, Spain

[3] Chemistry Division, Bhabha Atomic Research Centre, Trombay, Mumbai 400085, India

**Abstract:** We studied the effects of high-pressure on the crystalline structure of bulk and nanocrystalline scheelite-type PbMoO$_4$. We found that in both cases the compressibility of the materials is highly non-isotropic, being the *c*-axis the most compressible one. We also observed that the volume compressibility of nanocrystals becomes higher that the bulk one at 5 GPa. In addition, at 10.7(8) GPa we observed the onset of an structural phase transition in bulk PbMoO$_4$. The high-pressure phase has a monoclinic structure similar to M-fergusonite. The transition is reversible and not volume change is detected between the low- and high-pressure phases. No additional structural changes or evidence of decomposition are found up to 21.1 GPa. In contrast nanocrystalline PbMoO$_4$ remains in the scheelite structure at least up to 16.1 GPa. Finally, the equation of state for bulk and nanocrystalline PbMoO$_4$ are also determined.

---

[†] electronic mail: daniel.errandonea@uv.es, Tel.: (34) 96 354 4475, FAX: (34) 96 3543146
[*] Member of MALTA Consolider Team

**I. Introduction**

Lead molybdate (PbMoO$_4$), the mineral wulfenite, is well known as a superior medium for acousto-optic devices [1]. This material also attracts attention because of its great potential to be used as an effective cryogenic detector for double-beta decay experiments [2]. Thus, its physical properties have been subject of extensive research [3]. PbMoO$_4$ crystallizes at ambient conditions in the tetragonal scheelite-type (wulfenite) structure; space group (SG): $I4_1/a$, No. 88, $Z = 4$ [4]. On it, each Mo atom is surrounded by four equivalent O sites in tetrahedral symmetry. On the other hand, each Pb atom shares corners with eight adjacent MoO$_4$ tetrahedra forming PbO$_8$ bisdisphenoids.

The high-pressure (HP) behaviour of scheelite-structured oxides has been extensively studied during the last years [5]. High-pressure research has proven to be an efficient tool to improve the understanding of the main physical properties of tungstates and molybdates [5-8]. In particular, it has been established that most of them undergo a pressure-driven transformation involving a symmetry reduction from SG $I4_1/a$ to SG $I2/a$. According with Landau theory, this transformation has been characterized as a second-order transition [9]. CaMoO$_4$ and SrMoO$_4$ are two of the compounds following this structural sequence [7, 8]. In contrast with these two compounds, contradictory results have been published for isomorphic PbMoO$_4$. HP Raman studies were first carried out by Nicol and Durana to 4 GPa using NaCl as pressure medium [10], but they found no evidence for any phase transition. Jarayaman *et al.* extended these studies beyond 11 GPa using a 4:1 metahnol-ethanol mixture as pressure medium [11]. They detected a phase transition to an unknown crystalline phase at 9.5 GPa. More recently, Raman studies were performed up to 26.5 GPa by Yu *et al.* under the same pressure medium [12]. However, their results are in contradiction with those previously published. These authors observed subsequent disappearances of Raman peaks that assigned to a proposed gradual amorphization, which, according with the authors, is completed at 12.5 GPa. This contradiction and the fact that Raman spectroscopy cannot provide direct information on the crystal structure suggest that HP x-ray diffraction studies are needed to better characterize the HP structural properties of $PbMoO_4$.

Here we report angle-dispersive x-ray diffraction (ADXRD) up to 21.1 GPa for bulk and nanocrystalline $PbMoO_4$. The experiments are performed under similar experimental conditions than precedent experiments. They allow the accurate determination of the structural sequence and compressibility of bulk $PbMoO_4$. The similarities and differences between the HP behaviour of the bulk and nanocrystalline materials will be also discussed.

**II. Experimental details**

In order to prepare bulk $PbMoO_4$ appropriate amounts of PbO and $MoO_3$ were thoroughly mixed using acetone as grinding media. The homogenous mixture was pressed in to a pellet and heated a at 650°C for 12 h. Colorless product obtained was characterized by powder x-ray diffraction (XRD) pattern recorded on Panalytical X-pert Pro diffractometer using Cu $K_α$ radiation. The tetragonal wulfenite structure is confirmed at this stage. In order to further confirm the complete reaction the pellet was rehomogenized and heated at 750°C for 12 h in pellet form. The final product was characterized by comparing the XRD pattern with earlier reported data (Powder diffraction file 44-1486). The nanocrystalline sample of $PbMoO_4$ was prepared by hydrothermal method. Stoichiometric amounts of lead acetate and ammonium molybdate were dissolved separately in distilled water and transferred to a teflon lined stainless steel autoclave. The autoclave was heated at 180°C for 24 h. and then cooled slowly to ambient temperature. The solid parts of the sample were filtered and washed thoroughly with distilled water and dried at 60°C in air. The dried powder was characterized by powder XRD data. The powder XRD pattern of the nanocrystalline sample appears exactly similar to that of the bulk except for the absence of some weak reflection and significantly broadened reflections. The excess broadening of the sample compared to the bulk sample is attributed to the nano nature of the sample. The particle size estimated from excess broadening using the Scherrer relation is 20-25 nm.

Powder ADXRD measurements on both bulk and nanocrystalline $PbMoO_4$ have been carried out with an Xcalibur diffractometer (Oxford Diffraction Limited). X-ray diffraction patterns were obtained on a 135 mm Atlas CCD detector placed at 110 mm from the sample using $K_{\alpha 1}: K_{\alpha 2}$ molybdenum radiation. The x-ray beam was collimated to a diameter of 300 μm. High-pressure measurements on $PbMoO_4$ powder were performed in a modified Merrill-Bassett diamond-anvil cell (DAC) up to 21.1 GPa. The diamond anvils used have 500-μm-size culets. The same set-up was used previously to successfully characterize the HP phases of $HfTiO_4$ and $ZrTiO_4$ in the same pressure range [13]. The $PbMoO_4$ powders were placed in a 150 μm-diameter hole drilled on a stain-steel gasket, previously pre-indented to a thickness of 50 μm. A 4:1 methanol-ethanol mixture was used as pressure medium. It is known that this medium could induce non-hydrostatic stresses beyond 10 GPa [14, 15]. However, we decided to use it to allow a direct comparison with previous studies performed under the same pressure medium. Ruby chips evenly distributed in the pressure chamber were used to measure the pressure by the ruby fluorescence method [16]. Exposure times for ADXRD measurements were typically of 2 hours. The DAC used for these experiments allows access to an angular range $2\theta = 25°$. An exposure on the starting materials, at room conditions, in a 0.3-mm glass capillary was obtained using the same installation. The observed intensities were integrated as a function of $2\theta$ in order to give conventional one-dimensional diffraction profiles. The CrysAlis software, version 171.33.55 (Oxford Diffraction Limited), was used for the data collection and the preliminary reduction of data. The indexing and refinement of the powder patterns were performed using the POWDERCELL [17] and FULLPROF [18] program packages.

**III. Results and Discussion**

The room-pressure (RP) diffraction patterns of the bulk and the nanocrystalline samples have been properly indexed with the wulfenite structure. The lattice parameters obtained are slightly lower than those previously published. For the bulk we obtained $a$ = 5.427(2) Å and $c$ = 12.092(5) Å [V = 356.1(3) Å$^3$ and axial ratio $c/a$ = 2.228(1)] and for the nanocrystal we obtained $a$ = 5.424(2) Å and $c$ = 12.096(5) Å [V = 355.8(3) Å$^3$ and $c/a$ = 2.230(1)]. Differences among the bulk and nanocrystal parameters are smaller than 0.1%. This result is different to what was observed in scheelite $CaWO_4$, where a considerable volume expansion and a crystal symmetrization occur in the nanocrystal [19].

In order to analyze the HP results only the data collected below $2\theta$ = 18.3º are considered, because of the appearance of stainless steel peaks of the gasket at higher angles. However, as can be seen in Fig. 1, in this angular region there are enough Bragg peaks to obtain the pressure evolution of the unit-cell parameters of the low-pressure phase. As we will show latter, this region of the diffraction pattern is also enough to propose a crystalline structure for the observed HP phase. In the HP experiments performed in the bulk material we found that up to 9.9 GPa all the Bragg peaks can be accounted with the low-pressure wulfenite structure. Only the (202) reflection is not seen, as usual, since its intensity is smaller that 0.25% of the main reflection. At 11.5 GPa two extra peaks emerge as shoulders of the (101) and (112) reflections of wulfenite, respectively. The peaks are more prominent at 13 GPa; at this pressure other reflections also appear in the diffraction pattern; e.g. a weak peak near 2θ = 5.8º. Additionally, a broadening is clearly detectable in the reflections. Strongest extra peaks are indicated by arrows in Fig. 1. The changes found in the diffraction patterns indicate the occurrence of a phase transition. We locate the onset of the transition between 9.9 and 11.5 GPa; i.e. at 10.7(8) GPa. The transition is reversible, as shown in Fig. 1, but a small hysteresis is observed, being the low-pressure phase completely recovered below 6 GPa. These results are in agreement with the results of Jayaraman *et al.* [11]. They show that scheelite-type structure is stable up to near 10 GPa where a phase transition takes place to another crystalline phase. The HP phase remains stable up to the highest pressure covered by our experiments (21.1 GPa). We did not find any evidence of additional phase transitions, amorphization or decomposition. Therefore, the Raman peaks disappearance reported by Yu *et al.* [12] below 9 GPa should be related to experimental limitations and not to structural changes. In particular, these experiments could be affected by the lack of pressure-transmitting medium due to defective DAC loading causing a highly non-hydrostatic pressure environment. The same can be said about the proposed amorphization that was previously detected at 12 GPa [12]. Our samples remain crystalline at least up to 21.1 GPa. This conclusion is also supported by recent experiments in $PbMoO_4$ [20]. In addition, usually amorphization in scheelite-type oxides does not take place up to pressures close to 50 GPa [21].

As we mention above, many scheelite-structured compounds (isomorphic to wulfenite) undergo a pressure-driven structural transition to the monoclinic M-fergusonite structure (SG: *I2/a*, No. 15, *Z* = 4) [5]. This structure is a monoclinic distorted version of scheelite obtained by a small distortion of the cation matrix and significant displacements of the anions [7]. In our case, a Le Bail analysis [22] of the ADXRD patterns of the HP phase of $PbMoO_4$ shows that they can be fitted by the monoclinic M-ferguson­ite structure. Other monoclinic structures have been also considered (e.g. $LaTaO_4$-type, $HgWO_4$-type, $BaWO_4$-II, α-$MnMoO_4$-type, and wolframite [23, 24]). However, the computed x-ray diffraction patterns for these structures considerably differ from the patterns measured beyond 9.9 GPa. Thus, we concluded that M-fergusonite is the most likely structure of the HP phase of $PbMoO_4$. This conclusion is supported by the changes previously observed at 9.5 GPa in the Raman spectrum and the optical absorption edge [11]. In particular the phonon gap observed in the HP phase between 400 and 780 cm$^{-1}$ is characteristic of the M-fergusonite structure [25]. Another typical feature of it is the existence of four high-frequency modes which are associated to stretching modes of the $MoO_4$ tetrahedra [25]. Regarding the absorption edge, at the transition there is a collapse of the band-gap energy [11] becoming the sample suddenly deep yellow. The same band-gap collapse was observed in $PbWO_4$ at the scheelite-fergusonite transition [26]. Also the pressure evolution of the band-gap energy in the HP phase of $PbMoO_4$ is similar to that of M-fergusonite $PbWO_4$ [26].

The unit-cell dimensions for the proposed M-fergusonite structure of $PbMoO_4$ are at 11.5 GPa: $a$ = 5.325(6) Å, $b$ = 11.449(9) Å, $c$ = 5.282(5) Å, $β$ = 90.9(7)º, and V = 321.6(9) Å$^3$. The obtained parameters imply that, within the experimental accuracy, no volume discontinuity is observable at the transition. By analogy with other scheelite-type oxides [23], we attributed the observed transition to small displacements of the Pb cations from their high-symmetry positions. This structural instability would bring about changes in the O positions and the consequently polyhedral distortion. The monoclinic distortion of HP $PbMoO_4$ continuously increases from 11.5 GPa to 21.1 GPa enhancing these atomic displacements. In particular, the β angle gradually increases with pressure, viz. from 90.9º at 11.5GPa to 92.7º at 21.1 GPa. In addition, the difference between the *b/a* and *b/c* axial ratios of the M-ferguosonite phase also increases upon compression. A result of these changes is the augment of the splitting and broadening of the Bragg peaks of M-ferguosonite $PbMoO_4$ (see Fig. 1).

Fig. 2 shows a selection of diffraction patterns collected under compression in nanocrystalline $PbMoO_4$. This is the first time that HP diffraction is performed in scheelite-structured oxides. At ambient pressure the diffraction pattern is very similar to that of the bulk material. Again the weak (202) reflection is absent and the intensities of the rest of the reflections are similar than in the bulk. The only difference is that, as usual, the peaks in the nanocrystal are broader than in the bulk. The full-width at half maximum (FWHM) of the main reflection of wulfenite is 0.29º for the bulk at ambient pressure and 0.35º for the nanocrystal. The FWHM for the bulk is close to the instrumental resolution. Applying the Scherrer formula a particle size of 20-25 nm is calculated for the nanocrystals [27].

For nanocrystalline $PbMoO_4$ the diffraction patterns have been collected from ambient pressure up to 16.1 GPa. In contrast with the bulk, all the patterns can be assigned to the wulfenite structure. There is not any new Bragg peak that can be detected in the pressure range covered by the experiments. The only change is the broadening of the diffraction peaks caused by the deterioration of the quasi-hydrostatic conditions of the experiments beyond 10 GPa [13, 14]. Consequently, nanocrystalline $PbMoO_4$ remains stable in its ambient pressure scheelite-type structure up to 16.1 GPa. This fact indicates that the maximum stability pressure of the ambient pressure phase is higher in the nanocrystal than in the bulk. This is in agreement with the behaviour of other oxides [28].

From the analysis of all the x-ray patterns, we obtained pressure behaviour of the lattice parameters of both bulk and nanocrystalline scheelite-type $PbMoO_4$. The pressure dependence of the lattice and volume are given in Figs. 3 and 4. As can be seen from Fig 3 and 4, the behaviour of the bulk and the nanocrystal does not show significant differences. Only beyond 5 GPa the nanocrystal is slightly more compressible than the bulk material. In Fig. 3 it can also be seen that in wulfenite, the $c$-axis is more compressible than the $a$-axis. This fact is reflected in the decrease of the $c/a$ ratio from 2.228 (2.230) at 1 bar to 2.167 (2.176) at 9.5 GPa in the bulk (nanocrystal). This behaviour is coherent with that of other scheelite-type oxides [5]. The axial compressibilities of bulk (nano) PbMoO$_4$, defined as $\kappa_a = -\partial \ln a/\partial P$ and $\kappa_c = -\partial \ln c/\partial P$, are: $\kappa_a = 3.29 \cdot 10^{-3}$ GPa$^{-1}$ and $\kappa_c = 6.41 \cdot 10^{-3}$ GPa$^{-1}$ ($\kappa_a = 3.41 \cdot 10^{-3}$ GPa$^{-1}$ and : $\kappa_c = 6.93 \cdot 10^{-3}$ GPa$^{-1}$). These values are comparable to those of similar oxides [29]. The distinctive anisotropic compression of PbMoO$_4$ ($\kappa_c/\kappa_a \approx 2$) is due to the fact that the MoO$_4$ tetrahedra are rigid units, which are directly aligned along the $a$-axis, but along the c-axis there is a Pb cation intercalated between two MoO$_4$ units. Indeed most of the volume contraction of PbMoO$_4$ is accounted for the reduction of the Pb-O bond distances. Regarding the phase transition, the changes of the lattice parameters observed at the transition are similar to those of isomorphic compounds [5]. Basically one of the two identical axes of tetragonal scheelite increases slightly its value after the transition to monoclinic M-fergusonite while the other decreases its value.

The compression data for the low-pressure phase of PbMoO$_4$ have been fitted by a third-order Birch–Murnaghan equation of states (EOS) [30] giving the zero-pressure volume $V_0 = 356.1(8)$ Å$^3$, bulk modulus $B_0 = 67(4)$ GPa, and its pressure derivative $B_0' = 8(2)$ GPa for bulk PbMoO$_4$. For the nanocrystal we obtained: $V_0 = 355.9(8)$ Å$^3$, $B_0 = 69(5)$ GPa, and $B_0' = 6(2)$ GPa. The value of the bulk modulus is in agreement with that obtained by Hazen *et al.* [31] – 64(2) GPa - from single-crystal x-ray diffraction experiments performed up to 4 GPa. It also comparable to the 71(10) GPa value reported by Ying-Xin *et al.* [20]. However, these authors reported a disproportionate high $B_0$' equal to 29(3). This value is not compatible either with our experiments or with those reported by Hazen [31]. In addition it contradicts the fact that in most minerals related to scheelite $B_0$' is constrained between 3 and 8 [32]. Finally, it has been shown that the bulk modulus of scheelite-type oxides can be estimated from the cation bond distances (Pb-O in our case) and the cation formal charge (Pb in our case) [5]. Following this model we calculated $B_0$ = 71(11) GPa, which also agrees with the present experiments. To conclude, we would like to mention that in the case on HP phase of bulk $PbMoO_4$, the EOS of the low-pressure phase accurately describes its pressure-volume evolution.

**IV. Summary**

We have studied the structural properties of bulk and nanocrystalline $PbMoO_4$ under high-pressure up to 21.1 GPa. We found that in both materials the compression of the low-pressure phase is highly anisotropic being the *c*-axis the most compressible one. In addition, in the bulk material compression beyond 9.9 GPa induces a phase transition from the low-pressure scheelite-type structure to a high-pressure M-fergusonite-type structure. The transition is reversible and involves small displacements of the Pb cations from their high-symmetry positions. No evidence of additional transitions, decomposition, or amorphization is found. Finally, the EOS for bulk and nanocrystalline $PbMoO_4$ are determined. The compressibility of both materials is similar becoming the nanocrystal more compressible than the bulk beyond 5 GPa. The zero-pressure bulk modulus for the bulk is 67(4) GPa and for the nanocrystal is 69(5) GPa. These values agree with previous estimations of this parameter.

**Acknowledgements:** Financial support from Spanish MICCIN (Grants No. MAT2007-65990-C03-01, CTQ2009-14596-C02-01, and CSD2007-00045) as well as from Comunidad de Madrid and European Social Found: S2009/PPQ-1551 4161893 (QUIMAPRES).

**Figure Captions**

**Figure 1:** Selection of RT ADXRD data of bulk PbMoO$_4$ at different pressures up to 21GPa. In all diagrams, the background was subtracted. Pressures are indicated in the plot. In the ADXRD patterns at room pressure (RP) and 13 GPa we show with ticks the positions of the Bragg reflections for the low-and high-pressure phase. The arrows indicate the peaks that are characteristic of the HP phase.

**Figure 2:** Selection of RT ADXRD data of nanocrystalline PbMoO$_4$ at different pressures up to 16.1 GPa. In all diagrams, the background was subtracted. Pressures are indicated in the plot. In the ADXRD patterns at room pressure (RP) we show with ticks the positions of the Bragg reflections.

**Figure 3:** Pressure dependence of the unit-cell parameters in bulk and nanocrystalline PbMoO$_4$.

**Figure 4:** Pressure dependence of the volume in bulk and nanocrystalline PbMoO$_4$. The solid (dashed) line is the EOS obtained for the bulk (nanocrystal).

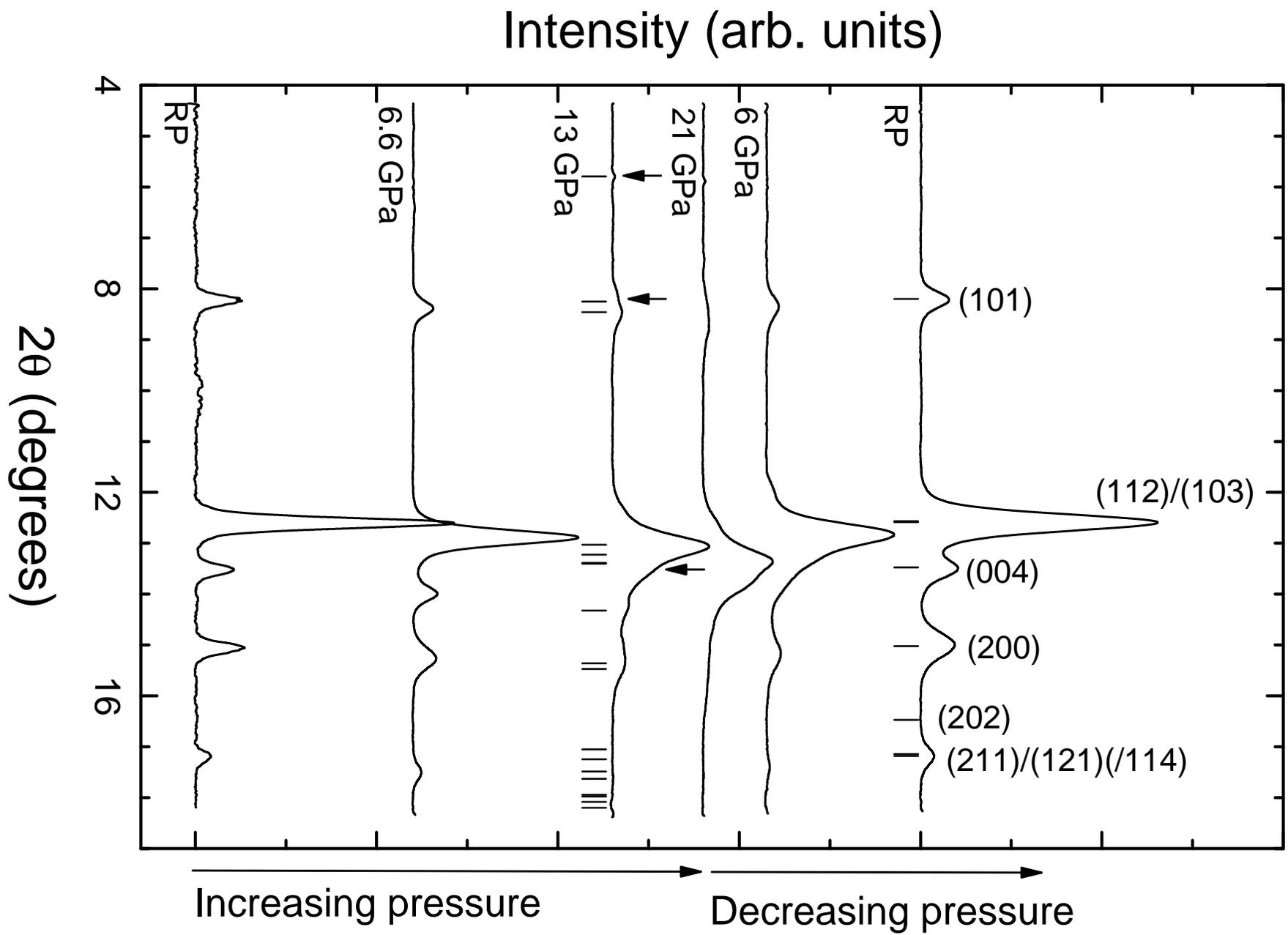

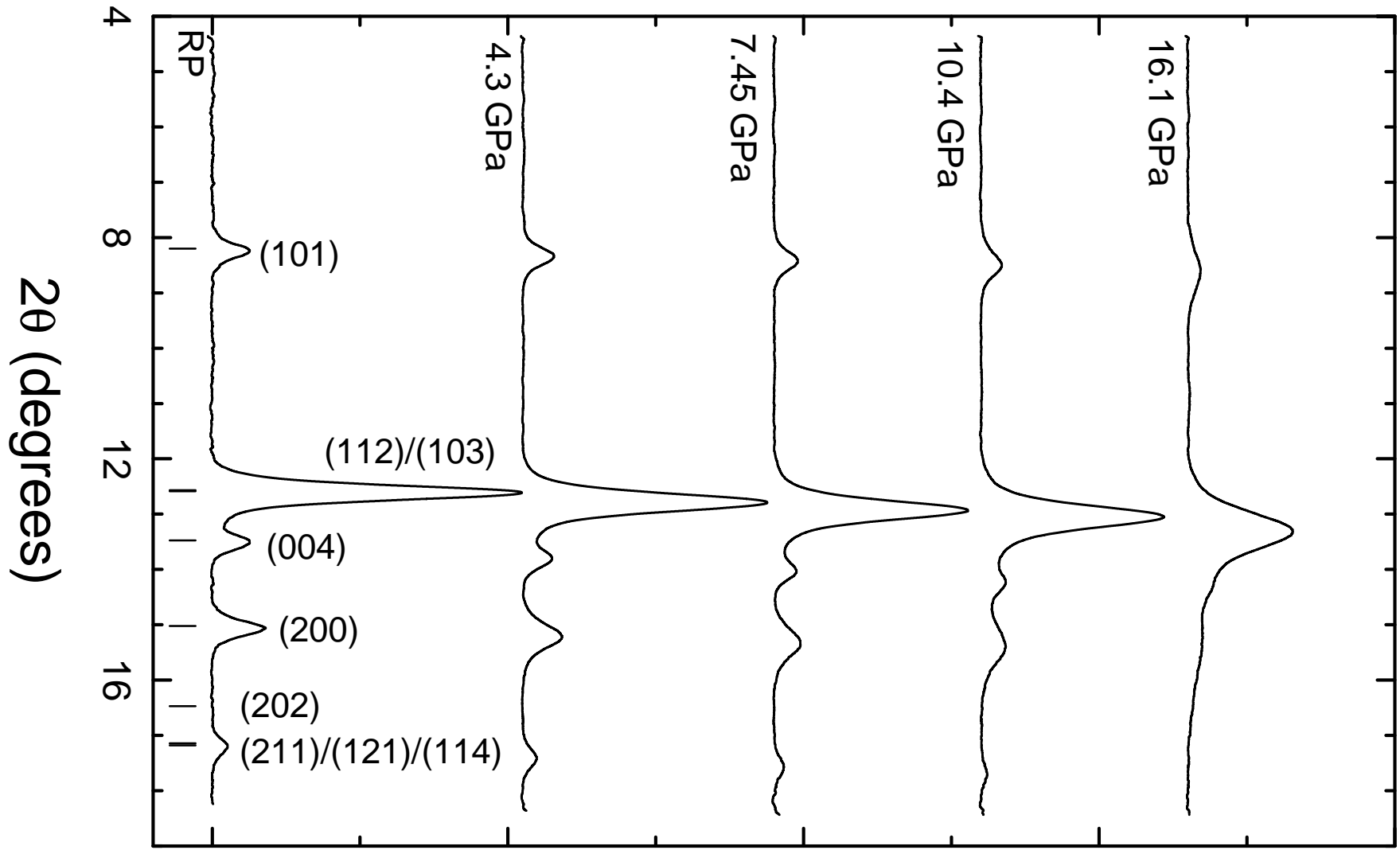

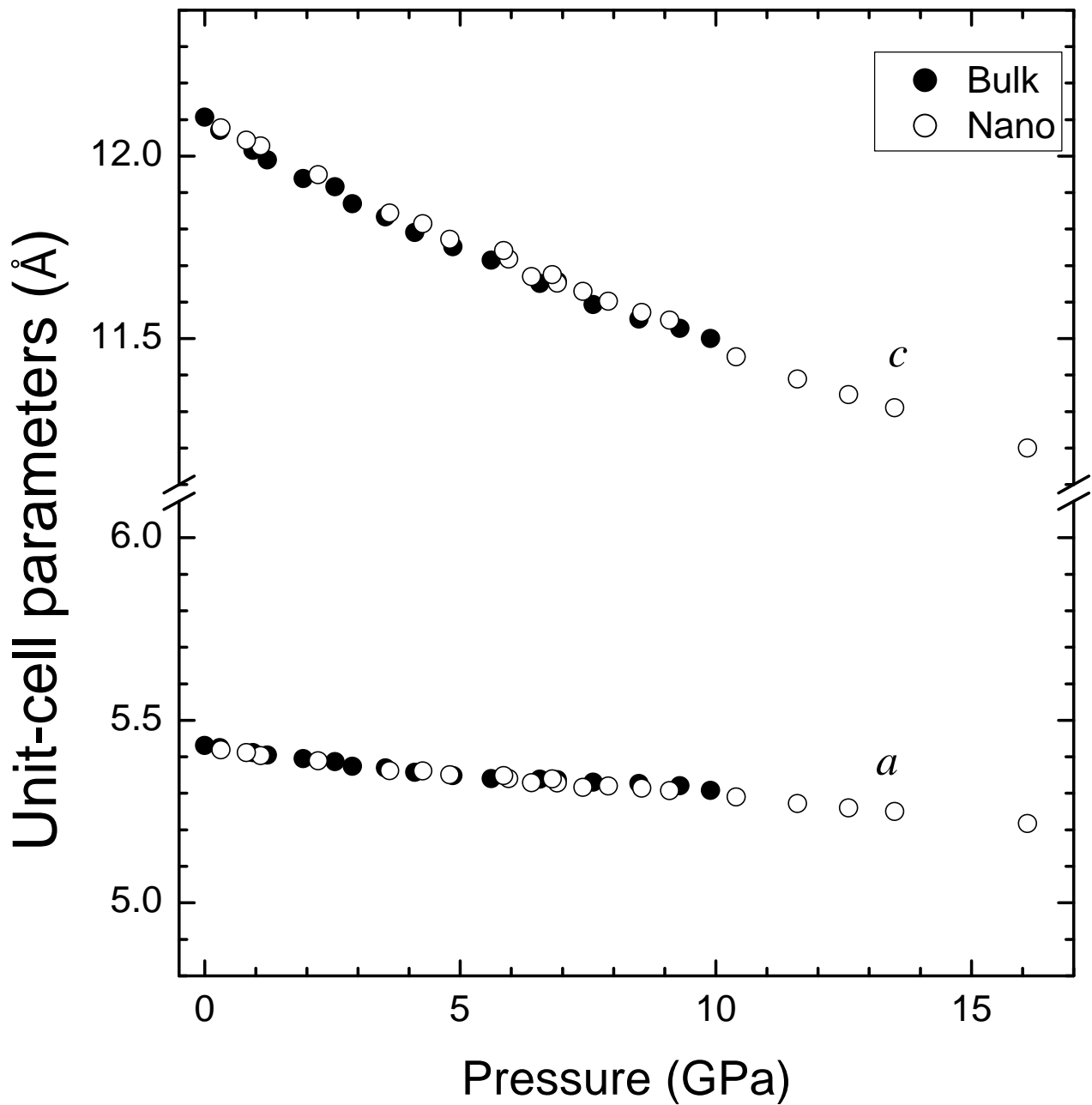

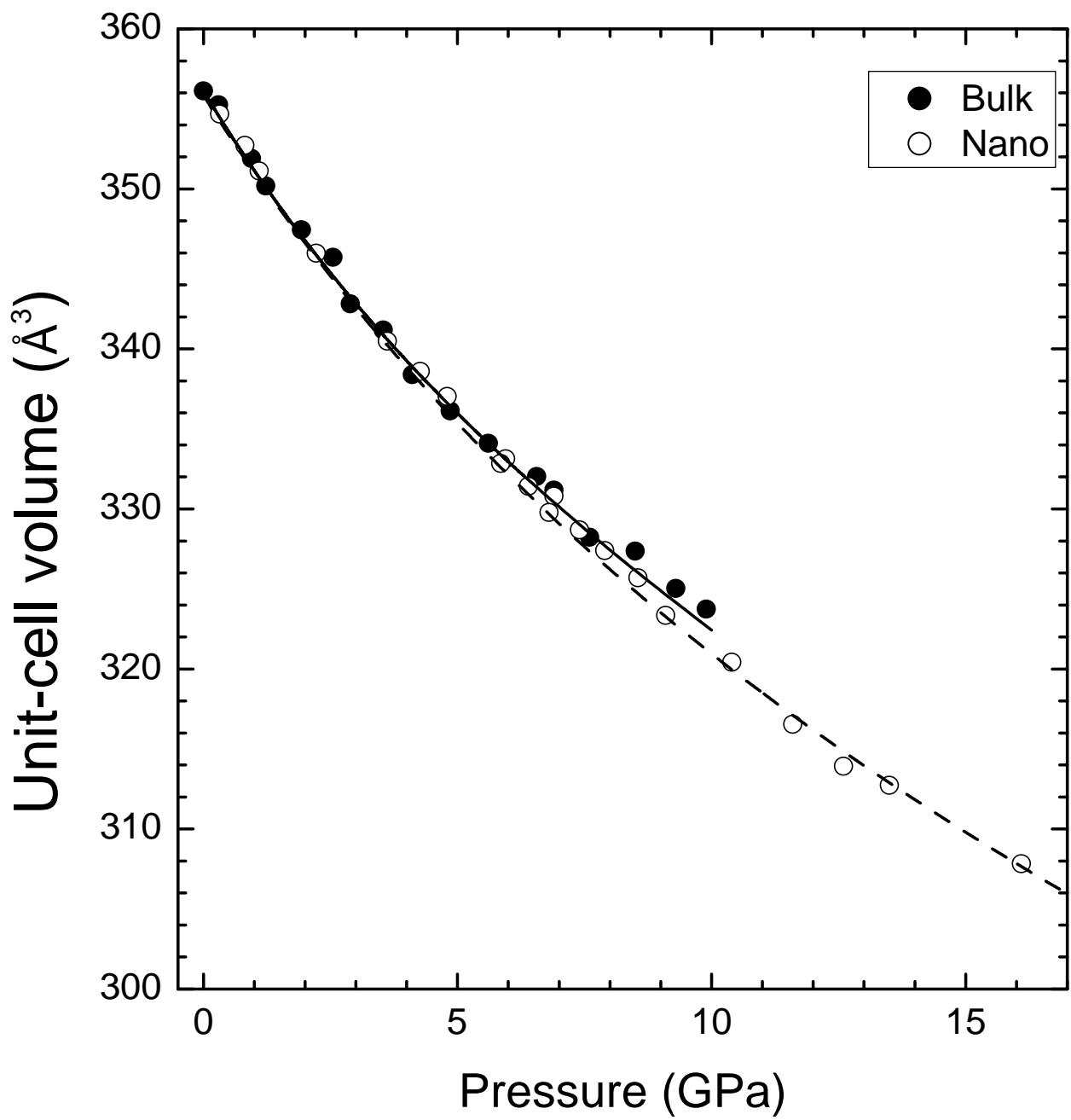